\documentclass{article}

\def\text#1{{\rm#1}}
\begin{document}

\author{Robert H. Swendsen, \\
Department of Physics, \\
Carnegie Mellon University, \\
Pittsburgh, PA 15213 \and Jian-Sheng Wang, \\
Department of Computational Science,\\
National University of Singapore, \\
Singapore 119260, \\
Republic of Singapore \and Shing-Te Li \\
IBM Corp., \\
Bldg. 334, M/S 2A1, \\
1580 Route 52, \\
Hopewell Junction, NY 12533 \and Brian Diggs, \\
Department of Physics, \\
Carnegie Mellon University, \\
Pittsburgh, PA 15213 \and Christopher Genovese, \\
Department of Statistics, \\
Carnegie Mellon University, \\
Pittsburgh, PA 15213 \and Joseph B. Kadane, \\
Department of Statistics, \\
Carnegie Mellon University, \\
Pittsburgh, PA 15213}
\title{Transition Matrix Monte Carlo }
\maketitle

\begin{abstract}
Although histogram methods have been extremely effective for analyzing data
from Monte Carlo simulations, they do have certain limitations, including
the range over which they are valid and the difficulties of combining data
from independent simulations. In this paper, we describe an complementary
approach to extracting information from Monte Carlo simulations that uses
the matrix of transition probabilities. Combining the Transition Matrix with
an N-fold way simulation technique produces an extremely flexible and
efficient approach to rather general Monte Carlo simulations.
\end{abstract}

\section{Introduction}

This paper presents both an approach to analyzing the information contained
in configurations generated by general Monte Carlo simulations, and a
closely related method of simulation that provides great flexibility and
surprising efficiency. The approach uses information contained in the
configurations about the set of possible changes on the next Monte Carlo
step, which we encode in a ``Transition Matrix.''\cite{TM time}$^{-}$\cite
{trans dyn estimators} This information is complementary to a histogram
analysis\cite{FS-histo}$^{,}$\cite{multi-can}. Indeed, it was originally
developed as an improvement to the histogram approach. The discovery that
the transition matrix, by itself, had definite advantages over the older
methods has opened up the development of several new techniques.

Combining the Transition Matrix with an N-fold way simulation technique\cite
{N-fold} produces an extremely flexible and efficient approach to rather
general Monte Carlo simulations. Because the same information is needed for
the simulation and the analysis, the two methods work extremely efficiently
together. In particular, the fact that this information is updated and used
for every step of the N-fold way simulation enables contributions to the
transition matrix to be made at every MC step, instead of after every sweep,
as in cluster simulations.

Since no use is made of cluster methods that could be limited to
non-frustrated systems, the techniques described in this paper are extremely
general. They can be used directly with any system that has equally spaced
energy levels, and, as is the case for histograms, can be generalized to
systems with continuous symmetry with the use of binning.

In the following sections, we will define the transition matrix, describe
how thermodynamic information is extracted from it, describe the N-fold way
(as it is used with the transition matrix), describe the creation of
generalized ensembles, and discuss the advantages of using these methods as
part of parallel simulations similar to Replica Monte Carlo or Parallel
Simulated Tempering.

\section{The Transition Matrix}

We will use the two-dimensional Ising model to demonstrate the transition
matrix approach, although generalizations to higher dimensions and more
complicated models is trivial. In particular, the generalization to a spin
glass is obvious. The Ising model is given by the Hamiltonian 
\begin{equation}
H=-J\sum_{\left\langle i,j\right\rangle }\sigma _{i}\sigma _{j}\text{,}
\label{hamiltonian}
\end{equation}
where $\sigma _{i}$ takes on the values $\pm 1$.

Although transition matrices can be defined for any kind of Monte Carlo
simulation that depends only of the energies of the initial and final
states, the primary transition matrix is defined in terms of the standard
Monte Carlo dynamics at infinite temperature. For this dynamics, each spin
is chosen with equal probability and ``flipped'' with probability one.

For every spin in a configuration, it is easy to calculate the change in
energy of the configuration when that spin is reversed by simply counting
the number of neighbors with the same sign. Denoting the number of sites
that will produce an energy change $\delta E$ by $n_{\delta E}$, we define
the ``infinite temperature transition matrix'' by 
\begin{equation}
T_{E,\delta E}\equiv \left\langle n_{\delta E}\right\rangle _{E}/N\text{,}
\label{TM}
\end{equation}
where $N$ is the total number of sites and the average is taken over all
configurations at energy  $E$.

The largest eigenvalue of this matrix is unity, and the corresponding
eigenvector is the density of states, $W(E)$. 
\begin{equation}
\sum_{\delta E}W(E-\delta E)T_{E-\delta E,\delta E}=W(E)  \label{eigen}
\end{equation}

The elements of the transition matrix must satisfy the condition of detailed
balance, 
\begin{equation}
W(E)T_{E,\delta E}=W(E+\delta E)T_{E+\delta E,-\delta E}\text{,}
\label{detailed balance}
\end{equation}
which requires the ``TTT-identity'', 
\begin{equation}
T_{E,\Delta }T_{E+\Delta ,\Delta }T_{E+2\Delta ,-2\Delta }=T_{E,2\Delta
}T_{E+2\Delta ,-\Delta }T_{E+\Delta ,-\Delta }\text{,}  \label{TTT}
\end{equation}
where $\Delta $ is the smallest allowable energy change.

For Potts models, the Ising model in three-dimensions, or other models with
a greater number of possible energy changes, there are additional
TTT-identities of the form, 
\begin{equation}
T_{E,(m-1)\Delta }T_{E+(m-1)\Delta ,\Delta }T_{E+m\Delta ,-m\Delta
}=T_{E,m\Delta }T_{E+m\Delta ,-\Delta }T_{E+(m-1)\Delta ,-(m-1)\Delta }\text{%
.}  \label{extend TTT}
\end{equation}
The situation is only slightly more complicated if other quantities, like
the magnetization or second-neighbor interactions, are introduced. In each
case, the introduction of an additional energy change creates two new
elements at each energy ($+\delta E$ and $-\delta E$), along with one new
identity (possibly involving four matrix elements), leaving one new
independent variable.

Transition matrices corresponding to general Monte Carlo dynamics, $\check{T}%
_{E,\delta E}$, can be constructed from the acceptance probabilities, 
\begin{equation}
\check{T}_{E,\delta E}=\left\{ 
\begin{array}{ll}
T_{E,\delta E}\times a_{E,\delta E} & \text{for  }\delta E\neq 0 \\ 
T_{E,\delta E}\times a_{E,\delta E}+\Sigma _{\delta E\neq 0}T_{E,\delta
E}\times \left( 1-a_{E,\delta E}\right)  & \text{for  }\delta E=0
\end{array}
\right. \text{.}  \label{ensembles}
\end{equation}
The corresponding probability distribution can then be constructed from the
leading eigenvector of this matrix.

For example, the standard Metropolis acceptance probabilities are given by 
\begin{equation}
a_{E,\delta E}=\min \left[ 1,\exp \left( -\delta E/kT\right) \right] \text{,}
\label{Metropolis}
\end{equation}
and the leading eigenvector of the corresponding transition matrix is the
usual canonical probability distribution.

\section{The N-Fold Way}

Instead of choosing spins and determining whether to accept each move, it is
possible to use the set of numbers, $\left\{ n_{\delta E}\right\} $, to
determine in advance what the probability of picking a spin that will change
the energy by $\delta E$ and calculating its probability of acceptance. A
class of spins with $\delta E$ is then picked with probability 
\begin{equation}
\left( \frac{n_{\delta E}}{N}\right) \left( \frac{a_{E,\delta E}}{A}\right) 
\text{,}  \label{Nf 1}
\end{equation}
where 
\begin{equation}
A=\sum_{\delta E}\left( \frac{n_{\delta E}}{N}\right) a_{E,\delta E}\text{.}
\label{Nf A}
\end{equation}
A spin from that class is chosen and flipped. The contributions to the
transition matrix (and any other quantity being computed) are weighted with
a factor of $\left( 1/A\right) $.

Since the set of numbers $\left\{ n_{\delta E}\right\} $ are updated at
every step for both the simulation and the transition matrix, little extra
computer time is needed to record contributions to the transition matrix at
every Monte Carlo step.

It is clear that any acceptance rates may be used in defining an ensemble
for Monte Carlo simulations, as long as detailed balance is satisfied.
Because of the reweighting of the contributions of each configuration, it is
not even necessary for the acceptance rates to be normalized.

\section{Generalized Ensembles}

To create an ensemble with a desired probability distribution, $P(E)$, a
necessary condition is 
\begin{equation}
P(E)\times T_{E,\delta E}\times a_{E,\delta E}=P(E+\delta E)\times
T_{E+\delta E,-\delta E}\times a_{E+\delta E,-\delta E}\text{.}
\label{multi balance}
\end{equation}
For example, the multicanonical ensemble\cite{multi-can} is characterized by 
$P(E)=P(E+\delta E)$, so that the condition on the acceptance rates is 
\begin{equation}
\frac{a_{E,\delta E}}{a_{E+\delta E,-\delta E}}=\frac{T_{E+\delta E,-\delta
E}}{T_{E,\delta E}}\text{.}  \label{multi accept}
\end{equation}
Eqn. \ref{detailed balance}  relates this to the usual condition that the
ratios of acceptance rates is equal to the ratio of the densities of state
at the two energies.

Eqn. \ref{multi accept} is not sufficient to determine the acceptance rates
uniquely. In addition to the usual acceptance rates given by the minimum of
the ratio in Eqn. \ref{multi accept} or unity, both 
\begin{equation}
a_{E,\delta E}=T_{E+\delta E,-\delta E}
\end{equation}
and 
\begin{equation}
a_{E,\delta E}=1/T_{E,\delta E}
\end{equation}
are valid. Many other options exist.

\section{The ``Equal-Hit'' Ensemble}

Although the multicanonical ensemble was designed to visit every energy
level with equal probability, that is not really optimal when using the
N-fold way. For low energies, the acceptance ratio is small and the N-fold
way achieves a uniform $P(E)$ by visiting an energy level very few times,
but weighting it with a large factor $\left\langle A^{-1}\right\rangle
_{E,Nf}$. To scan all energy levels equally, it would be more appropriate to
specify that the number of visits to each energy level, $H(E)$, is uniform.
Note that the average of $A^{-1}$ over the visits to energy level, $E$, with
the N-fold way is not the same as the microcanonical average over
configurations with that energy. Denoting the microcanonical average of a
quantity, $B$, by, $\left\langle B\right\rangle _{E}$, we have 
\[
\left\langle B\right\rangle _{E}=\frac{\left\langle BA^{-1}\right\rangle
_{E,Nf}}{\left\langle A^{-1}\right\rangle _{E,Nf}}\text{.} 
\]
If we take $B=A$, this gives 
\[
\left\langle A^{-1}\right\rangle _{E,Nf}=\frac{1}{\left\langle
A\right\rangle _{E}}\text{.} 
\]

Since the probability, $P(E)$, is given by the product of the number of
times an N-fold way move ends at the energy $E$ times the average inverse
acceptance ratio, $\left\langle A^{-1}\right\rangle _{E,Nf}$, the condition
for an equal-hit ensemble with $H(E)=H(E+\delta E)$ is 
\begin{equation}
\left\langle A^{-1}\right\rangle _{E,Nf}\times T_{E,\delta E}\times
a_{E,\delta E}=\left\langle A^{-1}\right\rangle _{E+\delta E}\times
T_{E+\delta E,-\delta E}\times a_{E+\delta E,-\delta E}\text{.}
\end{equation}

This gives the condition on the acceptance rates. 
\begin{equation}
\frac{a_{E,\delta E}}{a_{E+\delta E,-\delta E}}=\frac{\left\langle
A^{-1}\right\rangle _{E+\delta E,Nf}\times T_{E+\delta E,-\delta E}}{%
\left\langle A^{-1}\right\rangle _{E,Nf}\times T_{E,\delta E}}
\label{equal hit accept}
\end{equation}
This condition can also be fulfilled in many ways, including 
\begin{equation}
a_{E,\delta E}=\left\langle A^{-1}\right\rangle _{E+\delta E,Nf}\times
T_{E+\delta E,-\delta E}
\end{equation}
and 
\begin{equation}
a_{E,\delta E}=\left\langle A^{-1}\right\rangle _{E+\delta E,Nf}/T_{E,\delta
E}\text{.}
\end{equation}

\section{Equilibration}

At the beginning of a Monte Carlo simulation, there is usually little
information about either the density of states or the transition matrix. For
a canonical simulation at a given temperature, this does not cause a
problem. However, for either a multicanonical simulation or an equal-hit
simulation, such information is necessary. However, it turns out that the
problem of dealing with the lack of information is greatly simplified by the
transition matrices.

The primary approach consists of three stages.

For the first stage, we start with a random configuration and use one of the
acceptance rates if it is known, otherwise we arbitrarily set it equal to
unity. As the simulation in this stage progresses, we have a changing random
walk that does not satisfy detailed balance and is biased towards states
that have not yet been visited.

For the second stage, we take the approximate (and biased) transition matrix
from the first stage, and impose the TTT-identities. In the second stage of
the simulation, we use this approximate transition matrix to define the
acceptance rates. Although the acceptance rates are not those of the
ensemble that we really want to simulate, they do satisfy detailed balance
by their construction, so that the transition matrix calculated in the
second stage is not biased.

For the third stage, we use the estimate of the transition matrix (after
imposing the TTT-identities) from the second stage to determine the
acceptance rates. This algorithm represents a good approximation to the
ensemble we wish to simulate, as well as satisfying detailed balance. The
third stage is the main part of the simulation; it is where most of the
computer time is used.

A modification of the third stage would be to continually update the
transition matrix as the simulation progresses. This has the advantage that
the simulation provides an increasingly accurate representation of the
desired ensemble. The obvious disadvantage is that, strictly speaking, the
algorithm no longer satisfies detailed balance. However, we have carried out
some very long simulations on small systems with this modification, and
found that the errors decrease with the square root of the length of the
simulation, as expected. Note that the effect of any bias from the early
part of the simulation is expected to decrease directly as the length of the
simulation, rendering it unimportant. Therefore, we believe that the small
violation of detailed balance is harmless, and will improve the efficiency
of the algorithm.

The two-stage equilibration before the main simulation in the third stage
requires relatively little computer time, so that this approach turns out to
be more efficient than the standard multicanonical procedure, as well as
being much simpler.

\section{Parallel Simulations}

A great advantage of a multicanonical simulation, which is retained by the
equal-hit simulation, is that information over the entire range of energies
can be obtained. This means that a single simulation provides the
thermodynamic behavior at all temperatures -- even negative temperatures
(also known as the antiferromagnetic version of the model).

However, these advantages come with a price. The relaxation time to
equilibrate a random walk is proportional to the square of its range. Since
the total energy range is proportional to the number of particles, $N$, the
relaxation time in units of MC steps is proportional to $N^{2}$. This does,
indeed, turn out to be the case for the simulations described above.

We can greatly reduce this disadvantage by introducing parallel simulations.
If we consider a number, $l$, of parallel simulations on independent
replicas of the system, in which each replica is restricted to a range
proportional to $N/l$, the relaxation time for each replica is only
proportional to $\left( N/l\right) ^{2}$. Taking the increased number of
simulations into account, we are left with a relaxation time in units of MC
steps proportional to $N^{2}/l$. The relaxation time in units of MC
steps/site is proportional to $N/l$.

Note that this improvement occurs even with a serial machine. Implementation
on a parallel machine is trivial.

We can further improve the algorithm by exchanging replicas, as introduced
in the Replica Monte Carlo method\cite{Replica MC}$^{-}$\cite{high-T RMC}
and later rediscovered as parallel simulated tempering\cite{ParSimTemp}.

Break up the total energy range into pieces with equal width. Beginning with
the lowest energies, restrict the first replica to the first \textit{two}
pieces of the energy spectrum. The second replica is restricted to the
second and third pieces, and so on. Replica $n$ and replica $n+1$ therefore
overlap for half of their range of energies. After simulating for a few
relaxation times (proportional to $N^{2}/l$), all neighboring replicas that
are in the overlapping half of their ranges are exchanged. Since we are
either using a multicanonical simulation with single spin flips, or an
equal-hit ensemble with the N-fold way, the acceptance probability for such
exchanges is unity.

This combination of replica exchange with an equal-hit (or multicanonical)
simulation provides improvements over both. Although this approach has been
discussed in terms of transition matrix Monte Carlo, its use is much
broader. For example, it could easily be applied to multicanonical
simulations of biological molecules, as we intend to do in future work.

\end{document}